\documentclass{he_symp}
\usepackage{psfig,graphicx,epsfig}
\usepackage{color}
\usepackage{amsmath,amssymb,epic,eepic,array}
\begin{document}
\renewcommand{\FirstPageOfPaper }{ 155}\renewcommand{\LastPageOfPaper }{ 158}%%
%% This is he_symp_example.tex
%% LaTeX2e example style file for the 270. Heraeus Seminar on 
%% Neutron Stars, Pulsars and Supernova Remnants, held in Bad Honnef, Jan. 21-25, 2002 
%% Needs the LaTeX2e class file he_symp.cls
%%
% -----------------------------------------------------------------------------
%\documentclass{he_symp}
%\usepackage{psfig}
% -----------------------------------------------------------------------------
%\begin{document}

\title{On the Field-Aligned Particle Acceleration in the presence of Electron-Positron Pairs}
\author{S. Shibata\inst{1}, J. Miyazaki\inst{2},  \and F. Takahara\inst{2}}
\institute{Department of Physics, Yamagata University, 1-4-12 Kojirakawa, Yamagata 990, Japan 
\and Department of Earth and Space Science, Osaka University, 1-1 Machikaneyama, Toyonaka, Osaka 560-0043, Japan}
\authorrunning{S.Shibata et al.}
\titlerunning{Field-Aligned Particle Acceleration}

\maketitle

\begin{abstract}
We study self-consistency for the pulsar polar cap model.
In general, there will remain an unscreened electric field 
at the pair production 
front (PPF). We have derived the condition of electric field screening
by pairs in the presence of returning particles.
A previous belief that pairs with a density higher than the 
Goldreich-Julian density immediately
screen out the electric field beyond PPF is unjustified.
Pairs have little contribution on screening the electric field, and
the geometrical screening is the only known way of getting a finite 
potential drop. 
\end{abstract}

\section{Introduction}
One of the promising models for the pulsar action is the polar cap model
where the field-aligned electric field accelerates charged particles 
up to TeV energies, and causes
an electromagnetic shower. 
The potential drop is normally thought to be confined in a small region:
the field-aligned electric field is screened out at the both ends of
the accelerator (Fig.~1).
The polar cap potential drop is a part of the electromotive force
produced by the rotating magnetic neutron star. 
The voltage drop is a few percent of the available voltage for young pulsars,
while it becomes some important fraction for older pulsars.
The localized potential is maintained by a pair of anode and 
cathode regions, the formation mechanism of which is the long issue
of the polar cap accelerator.

\begin{figure}
\centerline{\psfig{file=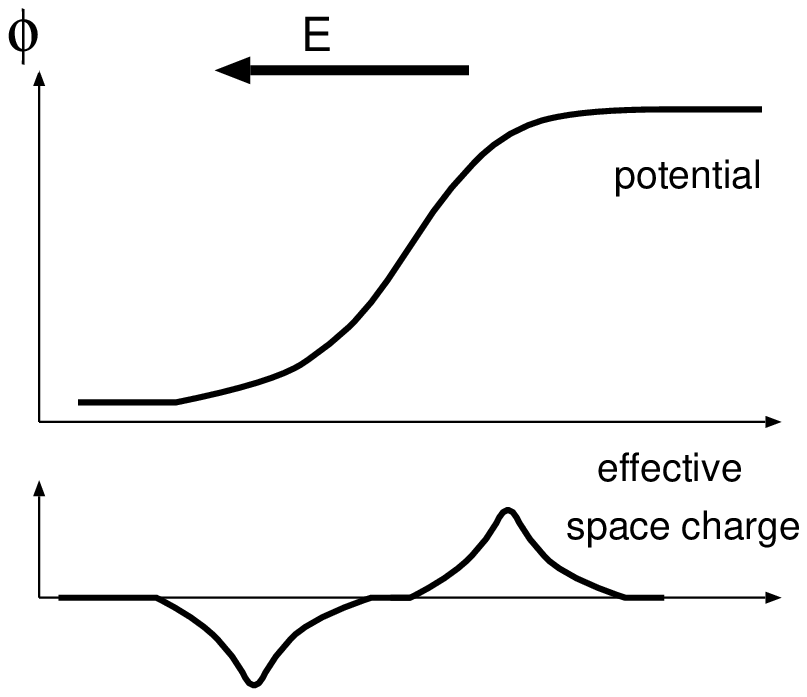,width=6cm,clip=} }
\centerline{\psfig{file=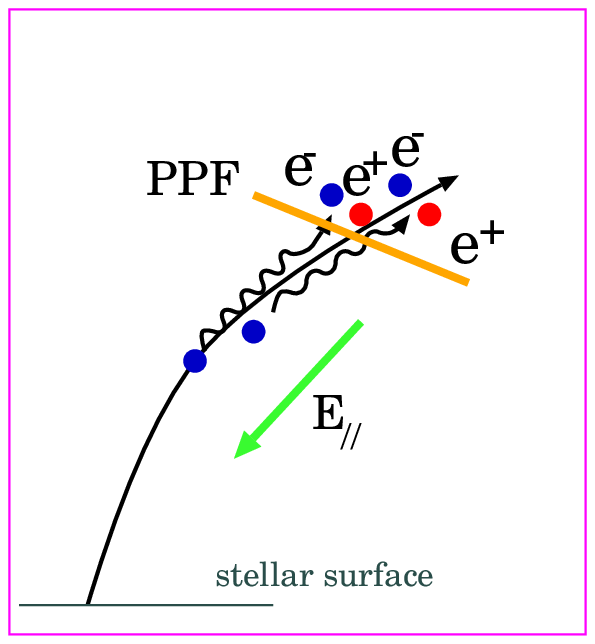,width=6cm,clip=} }
%\mbox{\includegraphics[width=6cm]{fig1.eps}}
%\mbox{\includegraphics[width=6cm]{cap.eps}}
\caption{{\small \bf localized potential drop}}
\end{figure}

\section{Pair Production Front (PPF) and Screening}

If the field-aligned electric field is formed to accelerate particles 
up to the Lorentz factor of $10^6 - 10^7$,
then a number of curvature gamma-photons convert into
electron-positron pairs under strong magnetic field 
$\sim 10^{12}$~G above a surface, which may be
called the pair production front (PPF).
The expected multiplicity --- number of pairs produced by one
primary particle --- is typically $10^3$ within
a distance $\sim 10^5 $~cm.

In a number of literatures, it is assumed that the field-aligned
electric field is screened out above PPF because the density
of pairs is higher than the Goldreich-Julian (GJ) density.
However, this screening process had not been studied in detail.
Shibata, Miyazaki \& Takahara (1998) showed that
pair polarization is not efficient even if pair density is much
higher than the GJ density.
In their work, presence of the particles reflected backward by unscreened
electric field was ignored.
In this paper, taking the backward-moving particles into account,
we derive an expression for the steady screening at PPF.

\section{Model}

Let us consider a model for the steady accelerator which
has a finite potential drop along the magnetic field
with the electric field screened at the both ends.
For definite sign of charge, let us assume electrons are accelerated
outwards, i.e., the electric field points toward the star.
$\gamma$-rays emitted by the electrons convert into pairs
beyond PPF continuously.

Schematic picture of the screening region is shown in Fig.~2.
At the top panel, electrons are accelerated to the right, so
the electric field points to the left. The gamma-rays emitted by
the electrons are converted in to pairs above PPF (the right side
of PPF in the figure).
Since the pairs are created almost continuously in a half space 
to the right of PPF, the pair-flux
monotonically increases with the coordinate $x$ along the magnetic
lines of force, where curvature is ignored since the scale considered here
is very short.

The second panel indicates the pair flux $F(x)$, which is defined as the
number of pairs created below $x$. The multiplication factor $m(x)$ may
be much convenient: $F(x)=m(x)j_0$ where $j_0$ is the primary flux of 
electrons coming up from a region below PPF.

Here, it must be reminded that why the accelerating electric field is
formed below PPF. 
The electric field well within the light cylinder is conveniently
decomposed by the co-rotational and non-corotational fields:
\begin{equation}
\vec{E}=- {\vec{\Omega} \times \vec{r} \over c} \times \vec{B}
- \vec{\nabla} \Phi,
\end{equation}
where $\Phi$ is the non-corotational electric potential, and field
aligned electric field is given by 
$E_\parallel = - \vec{B} \cdot \vec{\nabla} \phi / B$.
The Poisson equation for $\Phi$ reads
\begin{equation}
- \nabla^2 \phi = 4 \pi ( \rho - \rho_0 ),
\end{equation}
where $\rho_0 \approx \Omega B_z /2 \pi c$ is the GJ
charge density. 
From this expression it follows that 
the anode region is formed where $\rho > \rho_0$,
and the cathode region is formed where $\rho < \rho_0$.
In the screened region, $\rho = \rho_0$.

In our sign convention, 
the GJ density is negative, and electronic flow with super-GJ density
can produce cathode (negative) region, which is located above the
stellar surface (see Fig.~1).
There must be such a negative region below the PPF.
Above PPF, created pair positrons will be returned back by the electric field,
so they can destroy the cathode region. This fact sets a constraint to the
flux of pair positrons.
In the problem of electric field screening, one needs an anode region where
the effective charge $\rho - \rho_0$ is positive.
There are two possible ways of producing the anode region.
One is the `geometrical effect', and the other is pair polarization.

The geometrical effect is cased by the fact that the GJ charge density
is proportional to the magnetic component along the rotation axis $B_z$ while
the charge density of the electronic flow
is proportional to the magnetic field strength, or equivalently  inversely
proportional to the cross sectional area of the magnetic flux tube
if the flow velocity is $\approx c$ (constant).
Therefore, if the flow is curving toward the rotation axis,
the GJ density decreases slower than the primary electronic charge density
so that the electronic charge is insufficient to compensate the GJ density,
and the anode is formed. This effect is effective, and the accelerating
electric field can be terminated without pair creation.
Although geometrical screening is effective, PPF can in general appear
before geometrical screening. If this is the case, dynamics of pairs
just after PPF will be important.

Our interest in this paper is not the geometrical screening but the
pair effect. 
In general, at PPF, 
whatever values of the effective charge can be; it can
even be negative so that anode formation relay only on pair polarization.

As shown in Fig.~2, pair electrons are accelerated while
pair positrons are decelerated after their birth.
As far as the speed of pair positron is relativistic,
net space charge due to pairs are essentially zero. 
However, once the pair positron become non-relativistic,
positive space charge appears owing to charge conservation.
This is the effect which produces anode charge.
If the electric field is strong enough, some positrons are 
reflected backward and weaken the cathode space charge.
Assuming steadiness,
one can relate the strength of the electric field to the pair creation rate,
the returning positron flux,
and the effective space charge before PPF.
If the pair creation rate is more or less uniform, one can
do this analytically as shown in the following section.
Then, we have an screening condition.

\begin{figure*}
%\centerline{\psfig{file=shibata_fig2.eps,width=12cm,clip=} }
\centerline{\psfig{file=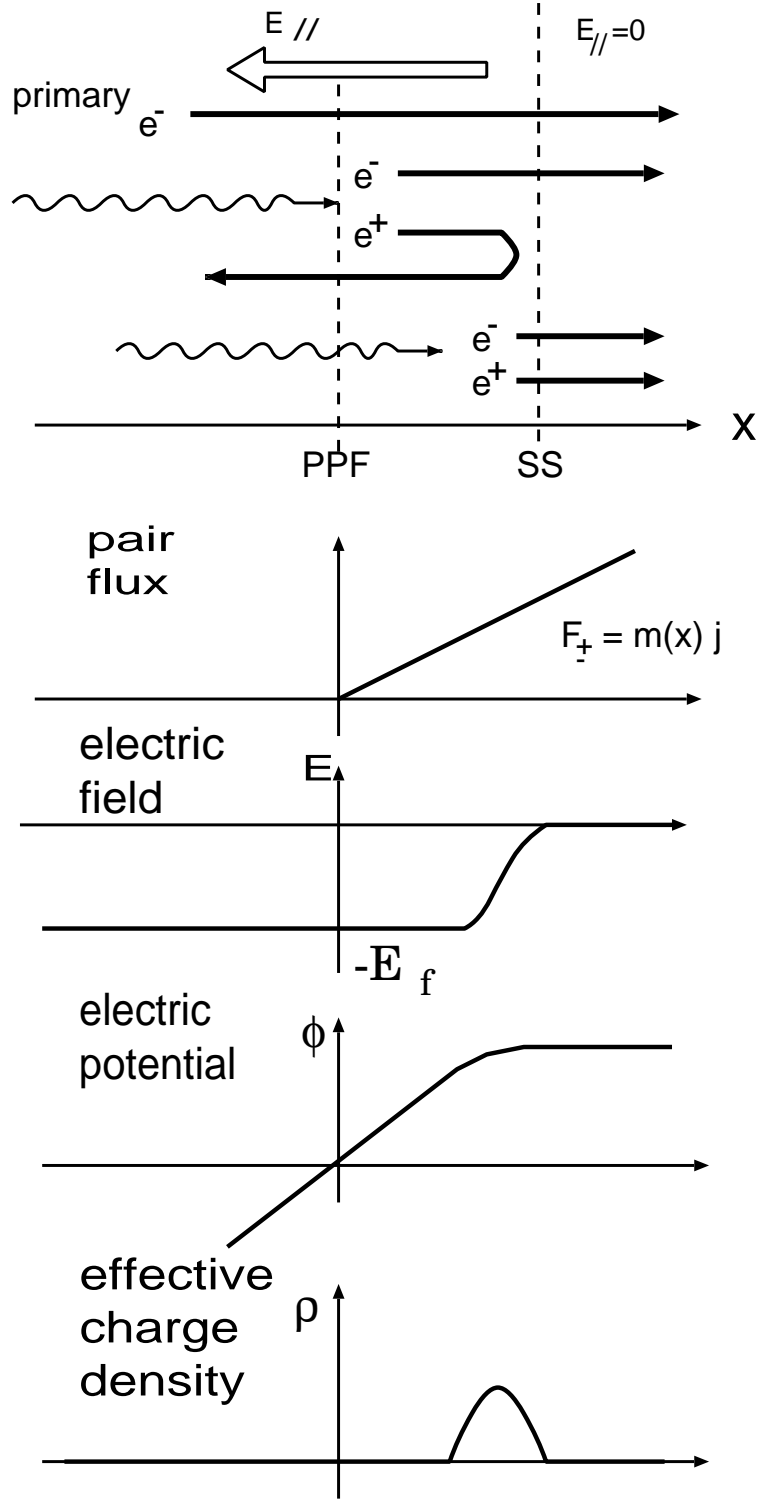,height=20cm,clip=} }

\caption{Schematic picture of screening region}
\end{figure*}

\section{Results}

If the pair creation rate is more or less uniform, the 
screening condition with returning positron can be derived 
analytically. Detailed calculation will be published near future, but
main results are  discussed below.

The field-aligned electric field $E_\parallel$ at PPF is screened
if the following two conditions are satisfied:
\begin{equation} \label{27}
{1 \over 2} E_{\rm f}^2 =
j \alpha_\phi [ \phi_1 (2-\phi_1) + \zeta ]
- (j-j_0-jM_1) \phi_{\rm s}.
\end{equation}
\begin{equation}
\rho_{\rm s} =
j \alpha_\phi (1-2 \phi_1) -(j-j_0-M_1j) > 0.
\label{e26}
\end{equation}
(\ref{27}) is just the result of Gauss' law: a condition for giving
the column positive charge to screen the electric field.
(\ref{e26}) is the positiveness of the space charge density at the screening
surface where $E_\parallel$ vanishes.

$E_{\rm f}$ is the normalized electric field to be screened:
$E_{\rm f}\equiv (e \lambda_0/mc^2)E_\parallel$, 
where $\lambda_0 = \sqrt{c^2/2 \Omega \Omega_B}=$ 
2.0 cm $B_{12}^{-1/2} P^{1/2}$ is the Debye length of
typical GJ plasma. For $10^{12}$ Volt acceleration in $10^4$ cm,
$E_{\rm f} \sim 100$.
$j$ is  
the primary electronic current density in units of the typical GJ density:
$j = J/(-\Omega B/2 \pi)$.
$\alpha_\phi$ is the pair multiplicity --- number of pairs produced
by one primary electron --- in an unit breaking distance
$\lambda_*$, with which
a non-relativistic particle turns by the unscreened electric field,
i.e., $e E_\parallel \lambda_* = mc^2$. 
In non-dimensional form, this distance is $1/E_{\rm f}$.
$\phi_1$ is defined in such a way that a pair positrons produced in between
$\phi=0$ (PPF) and $\phi_1$ are reflected backward, and
therefore
$M_1 = \alpha_\phi \phi_1$ gives the returning multiplication factor, i.e.,
$j M_1 = j \alpha_\phi \phi_1$ gives the returning positron flux.
$j_0$ is the normalized (local) GJ charge density: $j_0 = B_{\rm z}/B$.
$\zeta$: \ a numerical factor determined by the Lorentz factor 
at the birth of pairs; e.g., for $\gamma_{\rm birth} \sim 100$, $\zeta=1.7$.

\section{Application}

Positive space charge by pair polarization is little.
A previous belief that pair creation with a pair
density higher than the Goldreich-Julian density immediately
screen out the electric field is unjustified.
This is because \\
(1) only the non-relativistic positrons can
provide positive space charge, \\
(2) the non-relativistic positrons
are reflected backward by the electric field and leave the pair electrons
behind, so that a negative space charge is produced in the screening
region where the positive charge is required.

In the case where geometrical screening is ineffective,
more specifically, $j-j_0-jM_1 >0$, we have a condition
on the multiplication factor produced in unit breaking distance
$\lambda_*$ at PPF:
\begin{equation} \label{30}
\Delta M_{\rm screen} \ge
{ E_\parallel^2 \over 8 \pi mc^2 n_0 \zeta^\prime j },
\end{equation}
which indicates that if $j \sim 1$, a multiplication factor of $10^3$ 
is required in 0.01 cm. 
This condition will not be satisfied in the pulsar magnetosphere.

Thus the geometrical screening (toward curvature)
is the only known way of screening, i.e.,
in (\ref{27}), the positive effective charge $-j+j_0+jM_1 >0$
last in an appropriate distance (until the electric field is 
screened out or weakened enough to be screened by pair polarization).
Even on toward curved field lines, if the current density is super-GJ, or
if $E_\parallel$ at PPF is not so weakened, then
there is no way of maintaining a steady potential drop
along field lines.

Difficulty in pair screening is always true for field lines 
curving away from the rotation axis
regardless of the strength of the current density.

In our analysis, we do not take interaction between the particle beams
into account. Beam instability may produce some frictional force on
returning positrons and enlarge positive space charge. Time dependent
behavior such as repetition of acceleration and screening may be an alternative
way for a self-consistent model.

%\end{document}

\clearpage

\end{document}